\documentclass[superscriptaddress,aps,preprintnumbers,amsmath,showpacs,amssymb,prd,nofootinbib,reprint,twocolumn,floatfix]{revtex4-1}
\usepackage{bm, color}
\usepackage{amssymb,amsfonts,slashed,amsthm,amsmath,graphicx,soul,empheq}
\usepackage[caption=false]{subfig}
\usepackage{hyperref}
\usepackage{ulem} 

\begin{document}

\renewcommand{\figurename}{Fig.}
\renewcommand{\tablename}{Table.}
\newcommand{\Slash}[1]{{\ooalign{\hfil#1\hfil\crcr\raise.167ex\hbox{/}}}}
\newcommand{\bra}[1]{ \langle {#1} | }
\newcommand{\ket}[1]{ | {#1} \rangle }
\newcommand{\beq}{\begin{equation}}  \newcommand{\eeq}{\end{equation}}
\newcommand{\bef}{\begin{figure}}  \newcommand{\eef}{\end{figure}}
\newcommand{\bec}{\begin{center}}  \newcommand{\eec}{\end{center}}
\newcommand{\non}{\nonumber}  \newcommand{\eqn}[1]{\begin{equation} {#1}\end{equation}}
\newcommand{\laq}[1]{\label{eq:#1}}  
\newcommand{\dd}[1]{{d \o d{#1}}}
\newcommand{\Eq}[1]{Eq.~(\ref{eq:#1})}
\newcommand{\Eqs}[1]{Eqs.~(\ref{eq:#1})}
\newcommand{\eq}[1]{(\ref{eq:#1})}
\newcommand{\Sec}[1]{Sec.\ref{chap:#1}}
\newcommand{\ab}[1]{\left|{#1}\right|}
\newcommand{\vev}[1]{ \left\langle {#1} \right\rangle }
\newcommand{\bs}[1]{ {\boldsymbol {#1}} }
\newcommand{\lac}[1]{\label{chap:#1}}
\newcommand{\SU}[1]{{\rm SU{#1} } }
\newcommand{\SO}[1]{{\rm SO{#1}} }
\def\({\left(}
\def\){\right)}
\def\dt{{d \o dt}}
\def\diag{\mathop{\rm diag}\nolimits}
\def\Spin{\mathop{\rm Spin}}
\def\O{\mathcal{O}}
\def\U{\mathop{\rm U}}
\def\Sp{\mathop{\rm Sp}}
\def\SL{\mathop{\rm SL}}
\def\tr{\mathop{\rm tr}}
\def\ebq{\end{equation} \begin{equation}}
\newcommand{\OR}{~{\rm or}~}
\newcommand{\AND}{~{\rm and}~}
\newcommand{\EV}{ {\rm \, eV} }
\newcommand{\KEV}{ {\rm \, keV} }
\newcommand{\MEV}{ {\rm \, MeV} }
\newcommand{\GEV}{ {\rm \, GeV} }
\newcommand{\TEV}{ {\rm \, TeV} }
\def\o{\over}
\def\a{\alpha}
\def\b{\beta}
\def\c{\varepsilon}
\def\d{\delta}
\def\e{\epsilon}
\def\f{\phi}
\def\g{\gamma}
\def\h{\theta}
\def\k{\kappa}
\def\l{\lambda}
\def\m{\mu}
\def\n{\nu}
\def\p{\psi}
\def\q{\partial}
\def\r{\rho}
\def\s{\sigma}
\def\t{\tau}
\def\u{\upsilon}
\def\w{\omega}
\def\x{\xi}
\def\y{\eta}
\def\z{\zeta}
\def\D{\Delta}
\def\G{\Gamma}
\def\H{\Theta}
\def\L{\Lambda}
\def\F{\Phi}
\def\P{\Psi}
\def\S{\Sigma}
\def\me{\mathrm e}
\def\ol{\overline}
\def\tl{\tilde}
\def\*{\dagger}
\def\sout#1{}

\preprint{TU-1249}

\title{
A 
Bound on Light Dark Photon Dark Matter
}

\author{
Naoya Kitajima
}
\affiliation{Department of Physics, Tohoku University, 
Sendai, Miyagi 980-8578, Japan}

\author{
Shota Nakagawa
}
\affiliation{Tsung-Dao Lee Institute, Shanghai Jiao Tong University, \\
No.~1 Lisuo Road, Pudong New Area, Shanghai, 201210, China}
\affiliation{School of Physics and Astronomy, Shanghai Jiao Tong University, \\
800 Dongchuan Road, Shanghai 200240, China}

\author{
Fuminobu Takahashi
}
\affiliation{Department of Physics, Tohoku University, 
Sendai, Miyagi 980-8578, Japan}

\author{
Wen Yin
}
\affiliation{Department of Physics, Tokyo Metropolitan University, 
Minami-Osawa, Hachioji-shi, Tokyo 192-0397 Japan} 

\date{\today}

\begin{abstract}
We derive a 
bound on dark photon dark matter scenarios where the dark photon mass is generated through the Higgs mechanism, based on the requirement that symmetry breaking must occur sufficiently early in the universe. We emphasize that dark photon production occurs successfully when the dark Higgs field remains in the symmetric phase due to non-thermal trapping effects.
For renormalizable Higgs potentials, {our bound reads $$\frac{m_{\gamma'}}{q_H e_H}\;\gg \;60\,{\rm eV}\(\frac{2\pi}{\lambda}\)^{1/4}$$ where $m_{\gamma'}$ is the dark photon mass, $e_H$ is the gauge coupling,  $q_H$ is the charge of the dark Higgs boson, and $\lambda$ is the Higgs quartic coupling}. 
{This constraint holds independently of any complications arising from the Schwinger effect and vortex formation in the Higgsed phase.}
For more general Higgs potentials such as the Coleman-Weinberg type potential, our bound yields different forms.
 We argue that late-time symmetry breaking of the dark U(1) symmetry satisfying our bound has only a mild impact on both the abundance and momentum distribution of dark photon dark matter, and therefore does not pose any serious problem for the dark photon dark matter scenario.
\end{abstract}

\maketitle
\flushbottom

\section{Introduction 
\label{introduction}}
Dark matter accounts for about a quarter of the energy in the universe today, yet its fundamental nature remains elusive. This Letter focuses on dark photon dark matter. Along with axions, dark photon represents a promising candidate for light dark matter, attracting significant attention as a target for various experimental and theoretical studies.

Unlike axions, however, the conventional misalignment mechanism is known to be challenging for dark photons~\cite{Nakayama:2019rhg,Nakayama:2020rka,Kaneta:2023lki,Kitajima:2023fun,Fujita:2023axo} (see also Refs.~\cite{Nelson:2011sf,Arias:2012az}). Consequently, generating cold dark matter from very light dark photons is far from straightforward.
Several production mechanisms have been proposed, including {gravitational production during inflation~\cite{Graham:2015rva,Sato:2022jya,Cline:2024wja}} and reheating~\cite{Ema:2018ucl,Ahmed:2020fhc}, generation from axions~\cite{Agrawal:2018vin,Co:2018lka,Bastero-Gil:2018uel}, emission from cosmic strings~\cite{Long:2019lwl,Kitajima:2022lre}, resonant production from dark Higgs \cite{Dror:2018pdh,Harigaya:2019qnl} and production from spectator scalar field during inflation \cite{Nakai:2022dni}.  Thermal production with stimulated emission can produce dark photon around eV mass~\cite{Yin:2023jjj}. Non-thermal decay of heavy particle can produce dark photon by taking account of the stimulated emission~\cite{Moroi:2020has, Moroi:2020bkq, Nakayama:2021avl, Choi:2023jxw} (see \cite{Moroi:2020bkq} for the relation between narrow resonance and stimulated emission.)

{The dark photon mass can be generated through either the Higgs~\cite{Higgs:1964pj} or St\"uckelberg~\cite{Stueckelberg:1938hvi} mechanism.
In the Higgs mechanism case, the presence of dark photons can induce significant backreaction on the dynamics of the dark Higgs field. 
It was
pointed out in Refs.~\cite{Agrawal:2018vin,Kitajima:2021bjq} that the dark Higgs field acquires a large effective potential through its interaction with dark photons
\beq
\label{eff_mass}
\d V= - q_H^2 e_H^2 \vev{A_\m A^\m} |\f|^2,
\eeq
where $e_H$ is the gauge coupling, $q_H$ is the charge of the dark Higgs $\phi$,  $A_\m$  is the dark photon field, and the unitary gauge is assumed. Thus, the dark Higgs mass does not receive significant backreaction only if
\beq
\label{backreact}
-q_H^2 e_H^2 \vev{A_\m A^\m}  \ll m_h^2
\eeq
where $m_h$ is the dark Higgs mass at the minimum~\cite{Agrawal:2018vin}.
}

{
Focusing on dark photon production from axions,  we showed in Ref.~\cite{Kitajima:2021bjq} that the dark Higgs field can be trapped at the origin for a very long time due to this large effective mass (\ref{eff_mass}), enabling the dark Higgs to dominate the universe and generate significant entropy production. This behavior was also confirmed by
numerical lattice simulations. This mechanism bears similarities to the trapping of the flaton by thermal mass in thermal inflation \cite{Yamamoto:1985rd,Lyth:1995ka}. However, a significant distinction lies in the non-thermal nature of the mass generation, which can achieve substantially larger effective masses compared to thermal masses. 
Furthermore, in Ref.~\cite{Nakagawa:2022knn}, we demonstrated that this same mechanism can realize an early dark energy model capable of resolving the Hubble tension.}

In the scenarios of Refs.~\cite{Kitajima:2021bjq,Nakagawa:2022knn}, the dark Higgs is assumed to be in the symmetric phase when dark photons begin to be produced. This assumption stems from an important point emphasized in these references: dark Higgs particles would be produced via the Schwinger effect~\cite{Heisenberg:1936nmg,Schwinger:1951nm} when the effective potential (\ref{eff_mass}) gives a significant contribution to the dark Higgs potential, making the dark Higgs relatively light. This Schwinger production of dark Higgs particles then impedes the tachyonic production of dark photons, posing a serious challenge for dark photon production especially in the Higgsed phase.

Similar to the Schwinger effect issue, another significant challenge has been identified in Ref.~\cite{East:2022rsi}: for $\lambda \gg e_H^2$, vortex formation in the Higgsed phase during superheated phase transitions could impede the efficient production of dark photon dark matter with sufficiently low momenta. This vortex formation occurs because, above the superheating threshold, local magnetic field fluctuations induce decreases in the dark Higgs field, triggering further magnetic field clustering. This creates a positive feedback loop that ultimately drives the dark Higgs field to zero locally. Based on this observation {for the case of $\lambda \gg e_H^2$,} Ref.~\cite{Cyncynates:2023zwj} derived stringent constraints on dark photon parameters by requiring 
the condition (\ref{backreact}).
See also Ref.~\cite{Cyncynates:2024yxm}, a very recent follow-up work by the same authors providing a more detailed analysis of this issue.

{
In this Letter, we derive a 
constraint on dark photon parameters by focusing on dark photon dark matter production in the symmetric phase, rather than in the Higgsed phase. Importantly, dark photon production in this phase naturally proceeds without being hampered by the Schwinger effect or vortex formation. In this case, for the production of cold dark photon dark matter, symmetry breaking must occur by the time of redshift $z \sim 10^6$, which provides a limit on the dark photon parameters. 
Note that a symmetric phase is typically assumed as the initial condition in the Higgs theory, particularly in scenarios involving thermal trapping or non-minimal couplings. 
Combined with previous studies on symmetry-broken initial conditions, our analysis establishes a more comprehensive 
bound on dark photon dark matter. 
This general constraint holds regardless of the initial symmetry state, demonstrating that dark photon dark matter becomes incompatible with cosmological evolution if the system is in a symmetric phase at $z < 10^6$.
We will also discuss the effect of late-time symmetry breaking on dark photon dark matter, and show that its impact is  mild and does not pose any serious problem for the dark photon dark matter scenario.
}
\vspace{-2mm}

\section{General Bound on Dark Photon Parameters}
In this Letter, we focus on scenarios where dark photons acquire mass through the Higgs mechanism. A distinctive feature of these scenarios, in contrast to mass generation through the St\"uckelberg mechanism \cite{Stueckelberg:1938hvi}, is the existence of a {symmetric} phase where the dark Higgs field remains at the origin. {In this section we derive a 
bound on dark photon dark matter, focusing on the dark photon production in the symmetric phase.
}

The production of dark photons studied in Refs.\cite{Kitajima:2021bjq, Nakagawa:2022knn} proceeds in the symmetric phase. In contrast, in the Higgsed phase, the production of dark Higgs through the Schwinger effect\cite{Kitajima:2021bjq} and vortex formation~\cite{East:2022rsi} pose significant challenges. However, these problems can be suppressed if the dark photon production begins when the dark Higgs is in the symmetric phase, as the dark Higgs mass becomes progressively heavier during dark photon production. See Refs.~\cite{Kitajima:2021bjq,Nakagawa:2022knn} for mechanisms that initially trap the dark Higgs field at the origin.

The central focus of this Letter is a critical consequence of this mechanism: the substantial non-thermal effective mass of the dark Higgs field significantly delays the mass acquisition of dark photon dark matter.

If the dark photon mass $m_{\gamma'}$ is very small, but the gauge coupling $e_H$ is not that small, the symmetry breaking scale $v \sim m_{\gamma'}/e_H$ will be much smaller than the typical energy scale of the Standard Model, and a UV completion is needed.
Although our discussion could potentially be generalized to various scenarios such as symmetry breaking via non-perturbative effects\footnote{In such cases, the dark photon would affect the fundamental degrees of freedom. For instance, in a technicolor-type model, dark quarks would be subject to this effect, making it highly non-trivial whether the desired symmetry breaking occurs.}, we focus on the Abelian Higgs model as a UV completion. The Lagrangian is given by
\beq
\mathcal{L}=(\mathcal{D}_\mu\phi)^*\mathcal{D}^\mu\phi-\frac{1}{4e_H^2}F_{\mu\nu}F^{\mu\nu}-V(\phi),
\eeq
where the covariant derivative is denoted by $\mathcal{D}_\mu = \partial_\mu-iq_H A_\mu$ with $q_H$ being the charge of the dark Higgs $\phi$. $A_\m$  is the dark photon gauge field and $F_{\m\n}$ the field strength.
As a concrete exmaple, we consider the potential  given by
\beq
\label{ren_Higgs}
V(\phi) = \frac{\lambda}{4}(|\phi|^2-v^2)^2,
\eeq
where $\lambda$ and $v$ are the coupling constant and the vacuum expectation value, respectively. The charge of the Higgs field $q_H$ becomes physically meaningful when we introduce other matter fields. {The case of more general Higgs potentials will be discussed later.}
The potential (\ref{ren_Higgs}) contains a {tachyonic mass at the origin}
\beq
V(\phi)\supset -M_\f^2|\f |^2, ~~~M_\f^2 \equiv \frac{\l }{2} v^2,
\eeq
{which plays an important role in our discussion.}
The dark Higgs boson mass at the minimum is given by $m_h^2= \l v^2$. In the vacuum, the dark photon acquires a mass,
\beq
m_{\g'}= \sqrt{2} q_H e_H v,
\eeq
from the Higgs mechanism.

Suppose that the dark photon is the dominant dark matter component. This requires that its abundance matches the observed value~\cite{Planck:2018vyg} $$\rho_{\g'} = \rho_{\rm DM} \simeq 9.7\times 10^{-30} \GEV^4  \left(\frac{1+z}{10^6}\right)^3$$  and that it becomes cold enough by the time of redshift $z \sim 10^6$.\footnote{This is a conservative value to ensure dark matter is non-relativistic, based on the Lyman-$\alpha$ data and Milky-way subhalo count (see Refs.~\cite{Sarkar:2014bca,Corasaniti:2016epp,Das:2020nwc} for late forming dark matter). Using more precise limits, which is
beyond our scope, would not alter our conclusions.} 
For non-relativistic dark photons, the energy density is approximately given by 
\beq
\laq{DMdensity}
\rho_{\g'} \simeq \frac{m_{\g'}^2}{2}\vev{A^2}
\eeq
for $z\lesssim 10^6$. Here we defined $\vev{A^2}\equiv -\vev{A_\m A^\m}$.
Note that $m_{\g'}$ should not vary significantly with time to maintain the equation of state of matter (see Refs.\,\cite{Daido:2017wwb,Daido:2017tbr} for an example of time-varying dark matter mass). This means that $\f$ should be fixed at its VEV
$
\vev{\f}=v
$ by the time of $z \sim 10^6$, as this value determines the dark photon mass.
These requirements must be satisfied for dark photon dark matter regardless of its production mechanism.

A key observation is that for a very light dark photon mass, the field value $\vev{A^2}\neq 0$ becomes extremely large, particularly in the early universe, which gives an effective positive mass squared to the dark Higgs field, $\delta V$, in Eq.~(\ref{eff_mass})~\cite{Agrawal:2018vin,Kitajima:2021bjq}.
{If this effective mass becomes large, the dark Higgs field would be trapped at the origin for a long time~\cite{Kitajima:2021bjq,Nakagawa:2022knn}. However, for dark photons to constitute dark matter, the dark Higgs field must already be fixed at its VEV by $z \sim 10^6$. Therefore, this effective mass term at the origin must be subdominant at $z \sim 10^6$,} 
\beq 
\label{const2}
 q_H^2 e_H^2\vev{A^2} \ll M_\f^2 ,
\eeq
{which implies}
\beq
 \frac{\rho_{\g'}}{v^2}  \ll 
 {M_\phi^2 =}
 \frac{\l}{2} v^2, \laq{boundpre}
\eeq 
{where we used \Eq{DMdensity}, and the equality is for the Higgs potential (\ref{ren_Higgs}).}
We note that, although the constraint \eq{boundpre} is parametrically the same as (\ref{backreact}) derived in Ref.~\cite{Agrawal:2018vin} (and later adopted in Ref.~\cite{Cyncynates:2023zwj} based on the vortex formation~\cite{East:2022rsi}), they are derived from different physical reasoning. We also note that when the potential does not have a $\phi^4$ form, e.g., a Coleman-Weinberg type potential, our inequality (\ref{const2}) yields a different constraint. This is because what appears in our constraint (\ref{const2}) is the tachyonic mass at the origin, not the Higgs mass at the minimum. {See Appendix \ref{CW}.} 
The numerical confirmation of this condition (\ref{const2}) is given in appendicies. 

If this inequality (\ref{const2}) is not satisfied, the Higgs VEV would depend on the dark photon energy density, contradicting the requirement for dark photon dark matter. 
This implies that the dark photon energy density must be much smaller than the height of the Higgs potential in the symmetric phase. 
Given the dark matter abundance that dark photon should explain,
we obtain
\beq 
\l v^4 \gg 2\times 10^{-29}\GEV^4  \(\frac{1+z}{10^6}\)^3.
\eeq 
This directly leads to\footnote{For a subdominant dark matter comprising a fraction $r$ of the total dark matter, the right-hand side is multiplied by $r^{1/4}$. For comparison, the detection rate of the dark photon in various experiments scales with $e_H^2 r$.}
\beq \laq{bound}
\boxed{
\frac{m_{\g'}}{ q_H e_H}\gg 60 \EV \(\frac{2\pi}{\l}\)^{1/4}\(\frac{1+z}{10^6}\)^{3/4}.}
\eeq 
The bound on the dark photon mass becomes more stringent for smaller $\l$. {In particular, it holds even for the case of $\lambda \ll e_H^2$, where the vortex formation arguments in Ref.~\cite{East:2022rsi} are not applied.} Here, we choose the largest value allowed by perturbative unitarity. This bound, which is our main result, must be satisfied regardless of the dark photon production mechanism in the early universe, as we have only used the requirements for dark photon dark matter in the late universe.

In Fig.~\ref{fig:bound}, we show the bounds on the kinetic mixing $\chi=ee_H/6\pi^2$ as a function of the mass $m_{\gamma'}$.\footnote{{Note that introducing much larger kinetic mixing than this is not natural. {Indeed, such large} kinetic mixing cannot be generated from a renormalizable and perturbative model in quantum field theory.}}
{
Here we show two cases: the Mexican-hat potential (\ref{ren_Higgs})  and the Coleman-Weinberg type potential (\ref{CWpotential}).
The yellow shaded region is excluded by our condition (\ref{eq:bound}) for the Mexican-hat potential, while the orange shaded region corresponds to the case of the Coleman-Weinberg potential which is discussed in Appendix \ref{CW}. The blue shaded region is excluded by various experiments and observations \cite{Caputo:2021eaa} (see also \cite{AxionLimits} for summary of the bounds).
}
{
We note that some of the constraints assume the St\"uckelberg mass, which is valid when the presence of the dark Higgs is negligible, as in the case of the small gauge coupling considered here.}
\begin{figure}[t!]
\centering
\includegraphics[width=8.5cm]{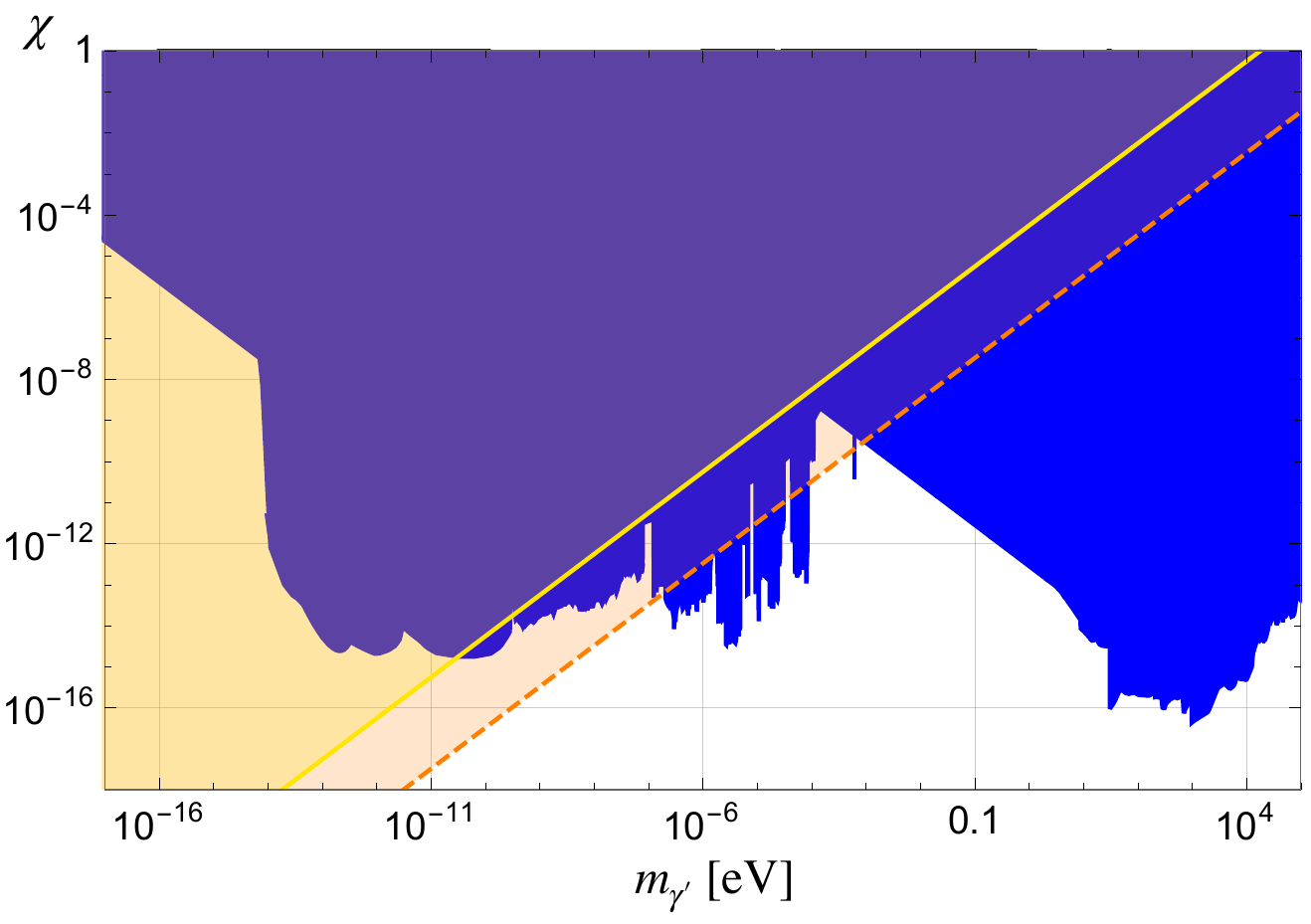} 
\caption{
The bound on kinetic mixing of dark photon dark matter as a function of the mass. 
{
Here we consider two cases: the Mexican-hat potential (\ref{ren_Higgs}) (yellow shaded region) and the Coleman-Weinberg type potential (\ref{CWpotential}) (orange shaded region).
Assuming the kinetic mixing $\chi =ee_H/6\pi^2$ with $e$ the electric charge, we choose $\lambda=2\pi$ and $q_H=1$ for the Mexican-hat potential, while we choose $\epsilon\equiv\sqrt{2}M_\phi/v=10^{-4}$ for the Coleman-Weinberg type potential.
Each shaded region is excluded due to our conditions (\ref{eq:bound}) and (\ref{CWbound}), while the blue shaded region is excluded by experiments and observations \cite{Caputo:2021eaa,AxionLimits}.
}
}
\label{fig:bound}
\end{figure}

{Before proceeding to the next section, let us discuss the assumption that the dark Higgs field is in the symmetric phase when the dark photon is produced.
There are several ways to initially trap the dark Higgs field. 
One possible approach is to introduce a nonminimal coupling to gravity. While this nonminimal coupling induces a Hubble mass term in the matter-dominated universe, it vanishes classically in the radiation-dominated universe. However, quantum effects can produce nonnegligible contributions that can trap the dark Higgs field in the symmetric phase with a sufficiently large coefficient.
{As alternative approaches}, we can consider {the thermal mass} of the dark Higgs. For instance, either introducing the Higgs portal coupling \cite{Patt:2006fw}, or assuming the inflaton decays slightly into the dark Higgs sector (establishing a dark thermal equilibrium)}\footnote{{If the dark Higgs mass is small, the dark temperature can be much lower than the photon temperature and we can neglect the dark thermal component in the late time cosmological history.}}, provides a thermal mass for the dark Higgs boson.
The specific parameter spaces where the dark Higgs field remains in the symmetric phase depend on the production mechanism of dark photons (see Ref.\cite{Nakagawa:2022knn} for the case of tachyonic production from axions). 

\section{Discussion and Conclusions}
In this Letter,
we have derived a constraint on light dark photon dark matter
whose mass is generated through the Higgs mechanism. 
The bound arises from requiring that the effective mass of the dark Higgs field induced by dark photons, first identified in Ref.~\cite{Kitajima:2021bjq}, must become subdominant by $z \sim 10^6$ for symmetry breaking to occur early enough to form cold dark photon dark matter. 
This leads to an interesting bound, $m_{\gamma'}/(e_H q_H)\gg 60{\rm eV} $, independent of specific production mechanisms. {In deriving the bound, we assume dark photon production in the symmetric phase, which naturally avoids the issues of Schwinger production of dark Higgs as well as vortex formation that arise in the Higgsed phase. This is one of the advantages of considering dark photon production in the symemtric phase.\footnote{Another one is that one can turn off the kinetic mixing with the ordinary photons during the production, which can also potentially hamper the efficient dark photon production~\cite{Kitajima:2021bjq}.} 
We also emphasize that our bound is not limited to the case of $\lambda \gg e_H^2$ where  the vortex formation arguments do not apply.}

The bound is particularly significant for ultralight dark photon dark matter scenarios, as it challenges models with small dark photon masses unless the gauge coupling is extremely tiny. 
Our work demonstrates that the dynamics of the dark Higgs field plays a crucial role in determining the viability of dark photon dark matter models.

When our derived inequality is satisfied, two distinct scenarios emerge: one where the dark Higgs field {remains in} the symmetric phase during dark photon production and another where it is in the broken phase throughout. 
In the latter case, there are issues of the Schwinger production of dark Higgs and the vortex formation, and the bound derived by requring the condition (\ref{backreact})  in Refs.~\cite{Cyncynates:2023zwj} is the same as our bound \Eq{bound}. 
As emphasized above, however, this agreement is only for the simple Higgs potential (\ref{ren_Higgs}), {and $\lambda\gg e_H^2$},  and for a more general Higgs potential such as the Coleman-Weinberg type potential, the two bounds are different. This is because it is the tachyonic mass of the dark Higgs at the origin that enters our bound (\ref{const2}). {See Appendix~\ref{CW} for a more detailed analysis.}

{Our analysis, in combination with previous studies on symmetry-broken initial conditions, establishes a stringent and comprehensive constraint on dark photon dark matter. This constraint remains valid regardless of the initial symmetry state, indicating that dark photon dark matter becomes fundamentally incompatible with cosmological evolution if the system maintains a symmetric phase at redshifts $z < 10^6$.}

{In the case where the dark Higgs field is trapped in the symmetric phase due to the large effective mass, the subsequent symmetry breaking would produce cosmic strings and dark Higgs condensate whose cosmological implications are worth investigating.
In contrast to the case of dark photon production in the Higgsed phase, however, we argue that this late-time phase transition has only a mild impact on the dark photon dark matter. This is because the energy of the dark Higgs condensate is comparable to that of dark photons immediately after the symmetry breaking occurs. Therefore, even if the decay of dark Higgs produces high-energy dark photons and these dark photons interact with each other through dark Higgs-mediated interactions, their momentum distribution becomes at most only marginally relativistic, and they soon become non-relativistic after a few Hubble times the number changing interaction such as $\g'\g'\g'\g'\to \g'\g'$ is efficient.
This potential warming-up of dark photons only makes our bound tighter, and in this sense our bound \Eq{bound} is conservative.
The decay of dark Higgs into dark photons and the subsequent thermalization processes are worth investigating further, and we leave their detailed study for future work.

\vspace{5mm}
{\bf Note added:} 
While we were preparing this manuscript, we became aware of Ref.~\cite{Cyncynates:2024yxm}, which appeared on arXiv and partially overlaps with our work. We emphasize that our bound is distinct, being derived from our previous finding of dark U(1) symmetry restoration caused by the non-thermal trapping of the dark Higgs field. Our analysis provides a novel constraint on dark photon dark matter {with the Higgs mechanism.}

\section*{Acknowledgments}
We thank Workshop on Cosmic Indicators of Dark Matter 2024 (15-17 Oct 2024),where the present work was initiated. This work is supported by JSPS Core-to-Core Program (grant number: JPJSCCA20200002) (F.T.), JSPS KAKENHI Grant Numbers 20H01894 (F.T.), 20H05851 (F.T. and W.Y.), 22H01215 (W.Y.), 22K14029 (W.Y.),  Incentive Research Fund for Young Researchers from Tokyo Metropolitan University (W.Y.). 
This article is based upon work from COST Action COSMIC WISPers CA21106, supported by COST (European Cooperation in Science and Technology).
Part of the results in this research were obtained using supercomputing resources at Cyberscience Center, Tohoku University.

\appendix

To supplement our main discussion, we show implications for dark matter production mechanisms and experimental searches, confirming the various important dynamics through lattice simulation.

\section{
Implications for Dark Matter Production and Detection
}

{So far, we have derived an interesting 
 limit on dark photon dark matter, showing that it disfavors the light dark photon mass range with much larger kinetic mixing than current experimental bounds. The limit in \eq{bound} also constrains cosmological scenarios for light dark photon dark matter, including dark matter production. We will explore some simple scenarios that satisfy \eq{bound} for production, though they require even smaller gauge couplings.}

Let us now discuss the implications of our bound \eq{bound}. We particularly focus on dark photon production with very light masses $m_{\g'}\ll \EV$, for instance, $m_{\g'} \lesssim 10^{-10}\EV$.
Dark photon production can occur through two distinct channels: production of transverse modes and production of longitudinal modes. For the transverse mode production, which proceeds through the gauge coupling, an efficient production mechanism requires a sizable gauge coupling $e_H$. For example, dark photon production through the tachyonic instability of an axion $a$~\cite{Agrawal:2018vin,Co:2018lka,Bastero-Gil:2018uel} requires a non-negligible axion-dark photon coupling $g_{a \g'\g'}\simeq \frac{e_H^2}{16\pi^2 f_a}$, where $f_a$ is the axion decay constant. Therefore, with small $e_H$ (corresponding to small $m_{\g'}$), efficient dark photon production becomes extremely challenging~\cite{Agrawal:2018vin}, {and a complicated clockwork set-up~\cite{Higaki:2016yqk} may be necessary.
{While production mechanisms for dark photon dark matter with a small gauge coupling have been proposed (see e.g., Refs.~\cite{Co:2021rhi,Kitajima:2023pby}), satisfying our derived constraints remains challenging, especially in the low mass regime. This highlights the substantial impact of our bounds on viable dark photon dark matter scenarios.}

The other possibility is to produce longitudinal modes of the dark photon. In this case, while the dark photon gauge coupling may be small, the production occurs through the Higgs coupling as the longitudinal mode corresponds to a would-be Nambu-Goldstone boson due to the equivalence theorem. However, as we will see, this scenario typically leads to very small gauge couplings.
Consider, for example, dark photons from cosmic strings. The correct abundance is achieved~\cite{Kitajima:2022lre} when $m_{\g'}\sim 10^{-13}\EV \(\frac{10^{14}\GEV}{v}\)^4$. While this satisfies our constraint, it requires an extremely small gauge coupling:
\beq
e_H q_H= 10^{-33}\left(\frac{m_{\g'}}{10^{-10}\EV}\right)^{5/4}.
\eeq

Another possibility is dark photon production through dark Higgs decay or late-time phase transitions (c.f.~\cite{Dror:2018pdh,Harigaya:2019qnl,Nakayama:2021avl}). We can parametrize the dark photon density as $\rho_{\g'}= c_1 \frac{\l}{4} v^4$ with $c_1< 1$, representing the fraction of Higgs energy converted to dark photons. The produced photon number density is $2\rho_{\g'}/m_h$. The initial dark photon momentum $p_{\g'}\simeq m_h/2$ satisfies $\frac{p_{\g'}}{m_{\g'}} \frac{1+z_{\rm nr}}{1+z_d}=1$, where $z_{\rm nr}$ and $z_d$ denote the redshifts at which the dark photon becomes non-relativistic and is produced through Higgs decay, respectively. For viable dark matter, we require $z_{d}>z_{\rm nr}>10^6$.
Due to redshift and $c_1<1$, the condition \eq{bound} (see \Eq{boundpre}) is naturally satisfied. However, the gauge coupling is constrained to be\footnote{Model building faces additional challenges that further restrict the parameter space, such as constraints on Higgs couplings for late-time phase transitions, suppression of production efficiency due to adiabaticity in phase transitions, and bounds on Higgs density~\cite{Nakayama:2021avl}.}
\beq
e_H q_H\lesssim 10^{-24}\(\frac{m_{\g'}}{10^{-10}\EV}\)^2.
\eeq
Such small gauge couplings make dark photon detection extremely challenging - they can be even weaker than gravitational coupling.
The detection of dark photon dark matter would, therefore, imply some non-trivial mechanism enhancing the interaction rate beyond the simplest and most natural scenarios {if the dark photon mass is generated by the Higgs mechanism.}

Another option is that the dark photon dark matter  is not very light. For instance, if the mass is heavier than the keV, it can easily avoid  our limit, and it could be probed in the conventional direct and indirect detection experiments~e.g. \cite{Alonso-Alvarez:2020cdv,Linden:2024fby}.

\section{Numerical simulation}
\lac{num}
Let us explore the non-thermal trapping numerically to strengthen our conclusions.  
The equations of motion for the Higgs and the gauge field are derived as follows
\begin{align}
\frac{1}{\sqrt{-g}} \mathcal{D}_\mu (\sqrt{-g} \mathcal{D}^\mu \phi) + \frac{\partial V}{\partial \phi^*}=0
\end{align}
\begin{align}
\frac{1}{\sqrt{-g}} \partial_\mu \left(\frac{1}{e_H^2} \sqrt{-g} F^{\mu\nu} \right)+2 {\rm Im}(\phi^* \mathcal{D}^{\nu} \phi) = 0.
\end{align}
In flat FLRW spacetime and imposing the temporal gauge condition, $A_0 = 0$, the above system of equations becomes
\begin{align}
&\phi''+2\mathcal{H}\phi'-\mathcal{D}_i \mathcal{D}_i \phi + a^2 \frac{\partial V}{\partial \phi^*} = 0 \\[2mm]
&E_i' + \partial_j F_{ij} +2 e_H^2a^2 {\rm Im}(\phi^* \mathcal{D}_i \phi) = 0 \\[2mm]
&\partial_i E_i + 2 e_H^2 a^2 {\rm Im} (\phi^* \phi') = 0,\label{eq:gausslaw}
\end{align}
where we use the conformal time ($\tau$) as time variable and the prime denote the derivative with respect to the conformal time, $a$ is the scale factor, $\mathcal{H} = a'/a$ and $E_i = F_{0j} = A_i'$. We assume the radiation dominated background, i.e., $a \propto \tau$ and $\mathcal{H} = 1/\tau$.

We perform numerical lattice simulations to solve the above equations of motion.
We set the initial field values by generating the random fields in Fourier space, satisfying
\begin{align}
\langle X(\bm{k}) X(\bm{k}')\rangle = (2\pi)^3 P(k) \delta^{(3)}(\bm{k} + \bm{k}'),
\end{align}
where $X = \mathrm{Re}(\phi),~\mathrm{Im}(\phi), A_i$ and the angular bracket represents the ensamble average. Here we adopt the white noise with cutoff scale $k_{\rm cut}$, i.e., $P(k) = \exp(-k/k_{\rm cut})$. We set $\phi'=0$ and $A_i'=0$ everywhere in coordinate space to satisfy the Gauss law (\ref{eq:gausslaw}).

\begin{figure}[t!]
\centering
\includegraphics[width=8.5cm]{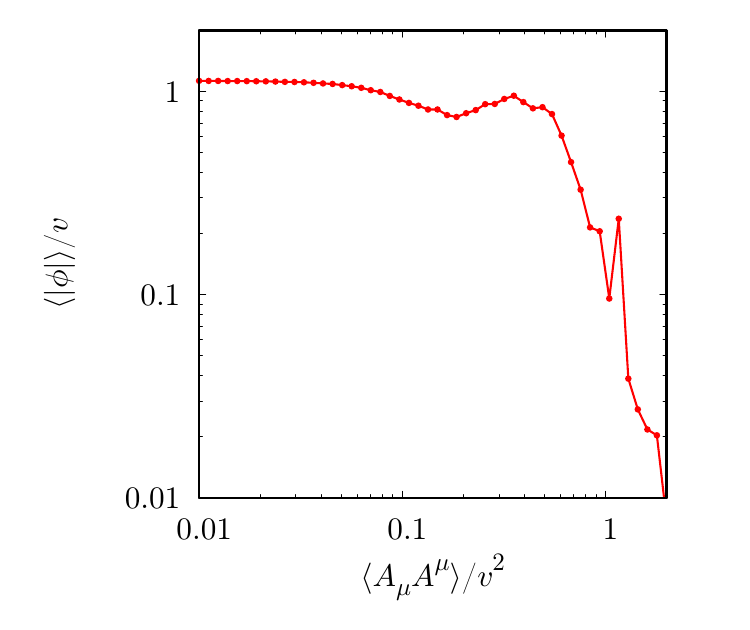}
\caption{
The relation between the field values of dark photons and dark Higgs.
Here we set $e_H = 1$, $\lambda = 2$.
}
\label{fig:phase_diagram}
\end{figure}

Fig.\ref{fig:phase_diagram} shows the relation between the spatial average (denoted by angular brackets) of the squared field values of dark photons at the beginning of the simulation ($v\tau = 1$) and that of the absolute field value of the dark Higgs at $v\tau = 5$. Here, we set $e_H = 1$ and $\lambda = 2$. As shown in this figure, the dark Higgs field begins to develop a nonzero expectation value when \Eq{boundpre} is saturated ($\vev{A^2} \simeq v^2$ in this simulation setup). This confirms that the symmetry breaking (or restoration) begins when \Eq{boundpre} is saturated\cite{Kitajima:2021bjq}.

\begin{figure}[t!]
\centering
\includegraphics[width=8.5cm]{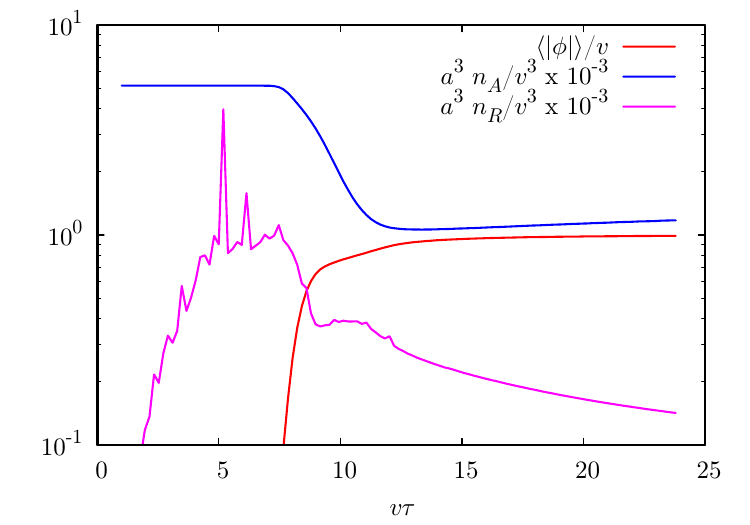}
\includegraphics[width=8.5cm]{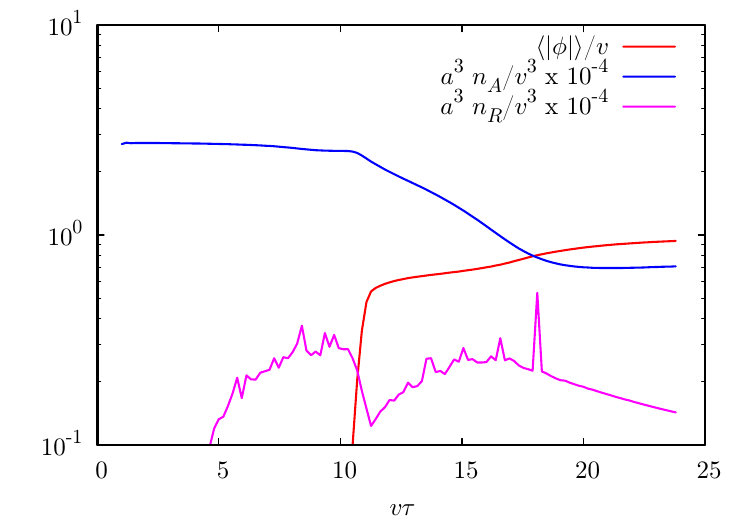}
\caption{
Evolution of $|\vev{\phi}|/v$ (red) and the comoving number densities of dark photons (blue) and dark Higgs (magenta).
Here we set $\lambda = 2$, $e_H = 0.2$, and the cutoff scale of the initial dark photon spectrum $k_{\rm cut} = H$ ($4H$) for the left (right) panel.
}
\label{fig:2}
\end{figure}
\clearpage
\begin{figure}[h!]
\includegraphics[width=7cm]{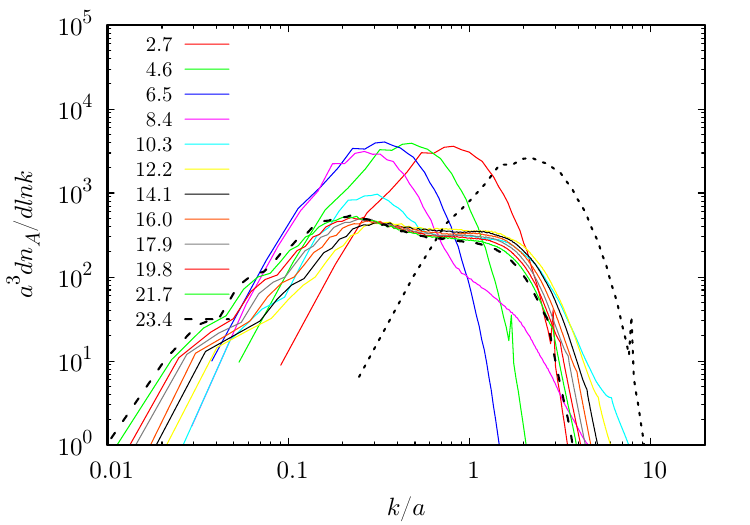}
\includegraphics[width=7cm]{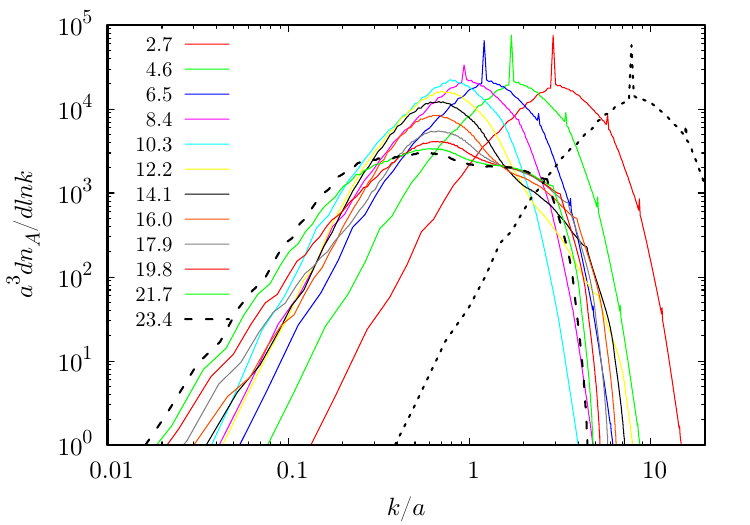}
\includegraphics[width=7cm]{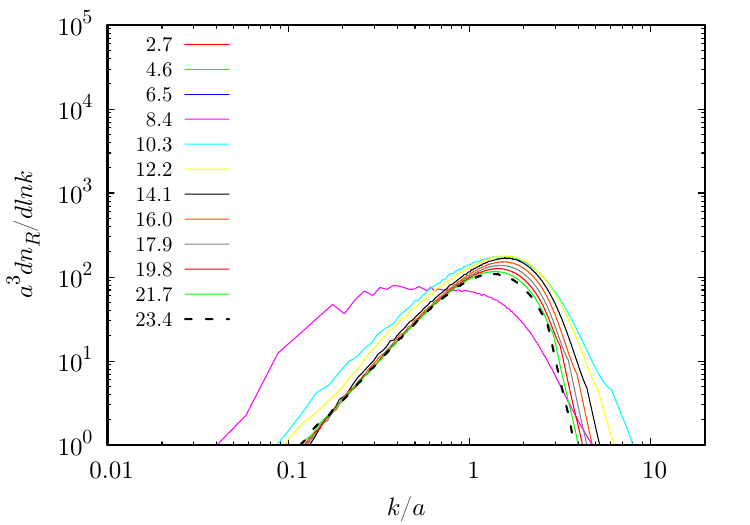}
\includegraphics[width=7cm]{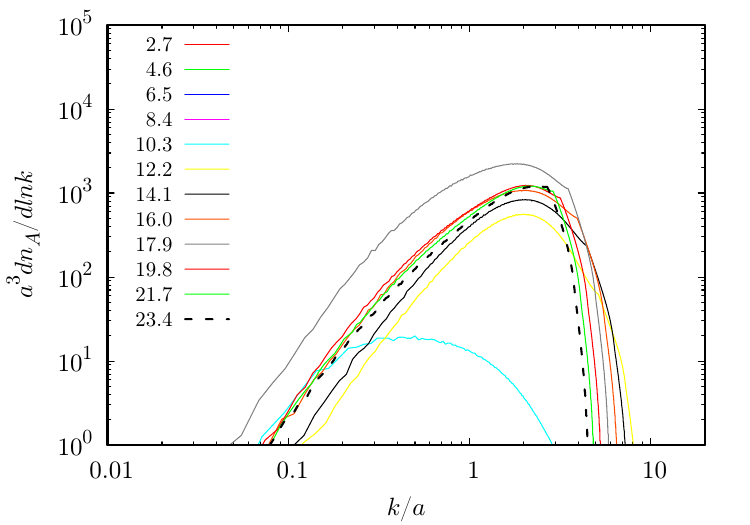}
\caption{
Time evolution of the spectrum of the comoving number densities of dark photons (top panels) and dark Higgs (bottom panels).
Here we set $\lambda = 2$, $e_H = 0.2$, and the cutoff scale of the initial dark photon spectrum $k_{\rm cut} = H$ ($4H$) for the left (right) panels. The dotted and dashed lines represent respectively the initial and final times.
}
\label{fig:3}
\end{figure}

Fig.\,\ref{fig:2} shows the evolution of $|\vev{\phi}|/v$ and the comoving number densities of the dark photon (including the Nambu-Goldstone mode) ($n_A$) and the dark Higgs (radial mode) ($n_R$) as a result of the lattice simulation with $512^3$ grid points. Here, we set the cutoff scale for the initial power spectrum to $k_{\rm cut} = H$ ($4H$) in the left (right) panel (see the dotted lines in the top panels in Fig.\,\ref{fig:3}). We find that the dark photons become non-relativistic and the comoving number density is fixed after the phase transition of the dark Higgs boson. Note that the dark Higgs production is not efficient, and the abundance of the dark Higgs is subdominant compared to that of the dark photons.

Fig.\,\ref{fig:3} shows the resultant spectra for dark photons (top panels) and dark Higgs (bottom panels) with $k_{\rm cut} = H$ ($4H$) in the left (right) panel. Before the phase transition, the spectrum of the dark photon is just red-shifted due to the cosmic expansion. After the phase transition, however, the peak is broadened to the high momentum region, which is due to the additional dark photon production from the Nambu-Goldstone mode. As shown in the figure, the original peak is marginally dominant over the newly produced peak in both the left and right figures. Thus, one can conclude that the cold dark photon dark matter production is successfully completed in this setup, insensitive to the initial momentum distribution of dark photons.

\section{{A case of Coleman-Weinberg potential}}
\label{CW}
{Let us consider the Coleman-Weinberg potential with a small tachyonic mass at the origin,
\begin{align}
    V(\phi) = - M_\phi^2 |\phi|^2 + \beta |\phi|^4  \left(\log\frac{|\phi|^2}{v^2} - \frac{1}{2}\right)
    \label{CWpotential}
\end{align}
with $M_\phi^2 \ll \beta v^2$. Then, the dark Higgs mass at the minimum $|\phi| \simeq v$ is given by $m_h^2 \simeq 2 \beta v^2$, which is much larger than the tachyonic mass at the origin, $M_\phi^2$. 
Then, while our condition (\ref{const2}) still holds, the bound on the dark photon mass and coupling becomes much tighter than \Eq{bound}: 
\begin{align}
\frac{m_{\g'}}{q_H e_H}\gg 10^3 \EV \(\frac{10^{-4}}{\epsilon}\)^{1/4}\(\frac{1+z}{10^6}\)^{3/4},\label{CWbound}
\end{align} 
where we have defined $\epsilon \equiv \sqrt{2} M_\phi/v \ll 1$. Note that, with this definition of $\epsilon$, one may simply replace $\lambda$ in \Eq{bound} with $\epsilon$ to obtain this result.

\bibliography{reference}

\end{document}